\documentclass[prb,twocolumn,showpacs,preprintnumbers,amsmath,amssymb,footinbib]{revtex4-1}
\usepackage{graphicx}
\usepackage[colorlinks=true,citecolor=red,urlcolor=blue]{hyperref}
\usepackage{dcolumn}
\usepackage{bm}
\usepackage{float}
\usepackage{color}
\usepackage{array}
\usepackage{longtable}
\usepackage{dcolumn}
\usepackage{braket}

\newcommand\mychi{\raisebox{0.35ex}{$\chi$}}

\begin{document}
\title{Deterministic Hadamard gate for microwave ``cat-state'' qubits in cQED}
\author{Simon E. Nigg}
\email{simon.nigg@unibas.ch}
\affiliation{Department of Physics, University of Basel,
  Klingelbergstrasse 82, CH-4056 Basel, Switzerland}
\begin{abstract}
We propose the implementation of a deterministic Hadamard gate for
logical photonic qubits encoded
in superpositions of coherent states of a harmonic oscillator. The proposed
scheme builds on a recently introduced set of
conditional operations in the strong dispersive regime of circuit
QED [Z.~Leghtas et al. Phys.~Rev.~A. {\bf 87} (2013)]. We further propose an architecture for coupling two such
logical qubits and provide a universal set of deterministic quantum
gates. Based on parameter values taken from the current state of the
art, we give estimates for the achievable gate fidelities accounting
for fundamental gate imperfections and finite coherence time due to photon loss.
\end{abstract}

\maketitle

\section{Introduction\label{sec:introduction}}
While digital (discrete) information encoding is classically advantageous over
analog (continuous) information encoding, this is no longer true in
the quantum case. Indeed, even when information is encoded into
eigenstates of operators with a discrete spectrum, the principle of
superposition allows for a continuum of errors. This has been recognized early
on and it was shown that continuous variable
quantum computation is theoretically as powerful as discrete
variable quantum computation~\cite{Lloyd-1999a}. Encoding
an effective two-level system into the large Hilbert space of a
continuous variable system, such as that of a harmonic
oscillator, may even be more advantageous by
allowing for more compact information processing and error correction
schemes~\cite{Gottesman-2000a}.

The last decade has witnessed steady improvements in
coherence times of quantum superconducting
circuits~\cite{Paik-2011a,Rigetti-2012a}, as well as significant advances in
quantum coherent frequency conversion between the optical and
microwave frequency ranges~\cite{Andrews-2013a}. These developments have revived interest in the possibility, originally proposed in the
context of linear optics~\cite{Ralph-2003b}, of encoding information in superpositions of
coherent states of light known as cat-states~\cite{Brune-1992a,Deleglise-2008}, in reference to
Schr\"odinger's famous thought
experiment~\cite{Schroedinger-1935}. The linear scaling of the decoherence
rate with the size of such states, in this case the
average number of photons, is an obvious drawback of such a
scheme. However, recently a proposal for encoding, manipulating and protecting
information in photon number parity eigenstates of microwaves
dispersively coupled to a superconducting qubit has been
put forward~\cite{Leghtas-2013a,Mirrahimi-2013a}. This provides an exciting new development in the
field of coherent state quantum computing.

In this work we follow an alternative route to implement a universal set of quantum
gates for coherent state qubits. In particular we provide a
fully deterministic Hadamard gate, the fidelity of which is
characterized for realistic parameters by solving the Lindblad master
equation numerically, accounting for the effects of photon loss in the dispersive
coupling regime. Probabilistic Hadamard
gates for coherent state qubits have been proposed in the linear
optics regime in Refs.~\onlinecite{Jeong-2002a,Marek-2010a,Ralph-2003b,Lund-2008a} and
implemented in Ref.~\onlinecite{Tipsmark-2011a}.

\section{System \& model\label{sec:system--model}}
\begin{figure}[ht]
\begin{center}
\includegraphics[width=0.9\columnwidth,clip]{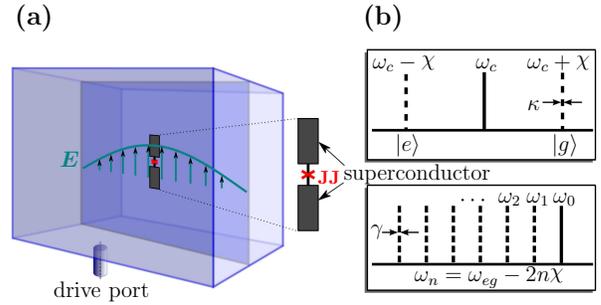}\caption{(Color
  online) {\bf
    (a)} A superconducting dipole antenna with a Josephson
  Junction (JJ) at its center (transmon) is coupled to the quantized electric field $\bm
  E$ inside a three dimensional
  microwave resonator. {\bf (b)} Schematics of the spectra of the
  cavity (upper panel) and transmon (lower panel) in the strong dispersive regime.~\label{fig:6}}
\end{center}
\end{figure}

The system we consider, called a 3D-transmon~\cite{Paik-2011a}, consists of a Josephson junction antenna, dipole-coupled to the
electric field of a microwave resonator as illustrated in
Fig.~\ref{fig:6}.
Approximating the non-linear
Josephson oscillator by a two-level system, consisting of its two
lowest eigenstates $\ket{g}$ and $\ket{e}$, we
model this system by the following Hamiltonian~\cite{Blais-2004a,Koch-2007a,Schuster-2007a}
\begin{equation}\label{eq:1}
{\bm H}_0=\omega_c{\bm a}^{\dagger}{\bm
  a}+\frac{\omega_{eg}}{2}{\bm\sigma}^z-\mychi{\bm a}^{\dagger}{\bm
  a}{\bm\sigma}^z.
\end{equation}
Here ${\bm a}$ annihilates a photon in the cavity mode with frequency $\omega_c$ and
${\bm\sigma}^z=\ket{e}\bra{e}-\ket{g}\bra{g}$ is a standard Pauli
operator for the two-level system with transition frequency $\omega_{eg}$. This model is valid in the dispersive (i.e. off-resonant) single mode coupling regime
$\mychi/\Delta\ll 1$ and for photon numbers satisfying
$n\ll n_{\rm crit}=\Delta/\mychi$, with
$\Delta=|\omega_c-\omega_{eg}|$. It is convenient to transform to a
rotating frame via the transformation
\begin{equation}
\bm U(t)=\exp\left[-i\left(\frac{\omega_{eg}}{2}{\bm\sigma}^z+(\omega_c+\mychi){\bm
  a}^{\dagger}{\bm a}\right)t\right].
\end{equation}
From now on we thus consider the Hamiltonian in this rotating frame
which reads
\begin{equation}\label{eq:12}
{\bm H}_0=-2\mychi{\bm a}^{\dagger}{\bm a}\ket{e}\bra{e}.
\end{equation}
In addition we shall consider the
following (classical) drive terms for the cavity ($\bm H_{\rm c}$) and the qubit
($\bm H_{\rm q}$), which in the rotating frame read
\begin{align}
\bm H_{\rm
  q}(t)&=\sum_{j=1}^{N_{0}}\frac{\Omega_j(t)}{2}\left(e^{i(\Delta^{(j)}_{\rm qd}t+\delta_j)}{\bm\sigma}^-+h.c.\right)\label{eq:5}\\
\bm H_{\rm
  c}(t)&=\sum_{j=1}^{N_{0}}\varepsilon_j(t)\left(e^{i(\Delta^{(j)}_{\rm cd}t+\mu_j)}{\bm
  a}+h.c.\right).\label{eq:6}
\end{align}
Here $\Delta_{\rm qd}^{(j)}=\nu_j-\omega_{eg}$ and $\Delta_{\rm
  cd}^{(j)}=\eta_j-\omega_c-\mychi$, where $\Omega_j$ and $\varepsilon_j$ are the drive strengths, $\nu_j$
and $\eta_j$ the drive frequencies and $\delta_j$ and $\mu_j$ the
drive phases. 

Starting from this model, it was
proposed in~\citet{Leghtas-2013a} and demonstrated experimentally
in~\citet{Vlastakis-2013a}, how to deterministically prepare superpositions of coherent
states of the cavity field, also known as cat-states. This is made
possible because although $\mychi$ is small compared with the
detuning between the cavity and the transmon, it is many orders of magnitude larger than the linewidths of the
transmon and cavity resonances, respectively given by $\gamma$ and
$\kappa$ (See Fig.~\ref{fig:6} (b)). This spectral resolution enables selective rotations of the transmon
conditioned on the number of photons in the cavity and selective
displacement operations of the cavity field conditioned on the state
of the transmon~\cite{Leghtas-2013a}.

In this work we account for the effects of photon loss in the Lindblad
master equation formalism. We shall assume that the transmon
is $T_1$-limited, i.e. that the intrinsic dephasing rate
$\gamma_{\varphi}=0$ and that the relaxation rate is solely due to the Purcell effect~\cite{Blais-2004a,Koch-2007a}, i.e.
that $\gamma=(\mychi/\Delta)\kappa$. The (zero-temperature) master equation for the
density matrix thus takes the form~\footnote{The Lamb-shift is
  implicitly
  absorbed into a renormalization of the cavity and qubit transition frequencies.}
\begin{equation}\label{eq:10}
\dot\rho = -i[\bm H,\rho]+\kappa\mathcal{D}[{\bm a}]\rho+\gamma\mathcal{D}[{\bm\sigma^-}]\rho,
\end{equation}
with $\mathcal{D}[{\bm A}]\rho=(2\bm A\rho\bm A^{\dagger}-{\bm
  A}^{\dagger}\bm A\rho - \rho{\bm A}^{\dagger}\bm A)/2$ and $\bm H=\bm
H_0+\bm H_{\rm q} +\bm H_{\rm c}$. This equation
is solved numerically using the Python library QuTip~\cite{Qutip-2012a}.

We follow the notation introduced
in Ref.~\onlinecite{Leghtas-2013a} and denote by $\bm D^{j}_{\alpha}$ a
displacement of the cavity field by the amplitude $\alpha$, conditioned
on the transmon being in state
$\ket{j}\in\{\ket{g},\ket{e}\}$. Similarly, a
rotation of the transmon by an angle $\theta$ around the axis
$\hat n_{\phi} =\cos(\phi)\hat x+\sin(\phi)\hat y$, conditioned on there being $n$
photons in the cavity is denoted with $\bm X^n_{\theta,\phi}$. Finally,
$\bm \Pi^e$ denotes the photon number parity operator conditioned on the
transmon being in the excited state $\ket{e}$. Ideally, these operations are given by
\begin{align}
\bm D^{j}_{\alpha}&=\exp(\alpha{\bm a}^{\dagger}-\alpha^*{\bm a})\otimes\ket{j}\bra{j}+\openone\otimes\ket{\overline
  j}\bra{\overline j}\label{eq:2}\\
\bm X^{n}_{\theta,\phi}&=\ket{n}\bra{n}\otimes
e^{i\frac{\theta}{2}\hat
  n_{\phi}\cdot\vec{\bm\sigma}}+\sum_{m\not=n}\ket{m}\bra{m}\otimes\openone\label{eq:3}\\
\bm\Pi^e&=\exp(i\pi{\bm a}^{\dagger}{\bm a})\otimes\ket{e}\bra{e}+\openone\otimes\ket{g}\bra{g},\label{eq:4}
\end{align}
where we used the notation $\overline g=e$, $\overline e=g$. Operations (\ref{eq:2}) and (\ref{eq:3}) can be realized by appropriately applying the drive terms
$\bm H_{\rm q}$ and
$\bm H_{\rm c}$ while operation (\ref{eq:4}) is realized by letting
the dispersive term~(\ref{eq:12}) act alone for a time $T_{\pi}=\pi/(2\mychi)$~\cite{Leghtas-2013a,Leghtas-2013b,Vlastakis-2013a}. Pulse imperfections and non-orthogonality of coherent states
with finite amplitudes result in small corrections as described in
more details in the supplementary material of Ref.~\onlinecite{Vlastakis-2013a}.
\section{Coherent state qubits\label{sec:coher-state-qubits}}
In binary quantum logic, the two computational states may be encoded in any two
mutually orthogonal states. Even in the case of discrete variable
quantum computation there thus exists a continuum of possible
encodings. In the discrete variable case, the computational states are
typically chosen to be the
ground and the first excited states of the system representing the
qubit. This choice is advantageous because these states are
stationary, i.e. their time evolution is given by a (trivial) phase
factor and relaxation provides a natural way to reset the qubit. Different types of encodings have been considered also for
continuous variables and it has been shown theoretically that universal
quantum computation is possible also in this case~\cite{Lloyd-1999a}. Of interest
here is the continuous variable encoding of information in superpositions of coherent states of a
harmonic oscillator, first
proposed in~\citet{Ralph-2003b} and~\citet{Gilchrist-2004a} in the context of linear
optics. Importantly, coherent states (with non-zero amplitude) are not eigenstates of the
harmonic oscillator Hamiltonian $\bm H_{\rm osc}=\omega{\bm a}^{\dagger}{\bm a}$, but under the action of the latter their phase evolves periodically
as $\ket{\alpha(t)}=\exp[-i\omega t{\bm a}^{\dagger}{\bm
  a}]\ket{\alpha(0)}=\ket{e^{-i\omega t}\alpha(0)}$; a property which
is made use of in the implementation of the photon number parity
gate $\bm\Pi^e$ given above. Coherent states are often
described as classical states of light owing to the fact that they are
steady state solutions of the damped, classically driven, harmonic
oscillator. The latter property lies at the heart of recent proposals
to stabilize superpositions of coherent states -- clearly
non-classical states of light -- by engineering an appropriate
dissipative environment~\cite{Everitt-2013a,Mirrahimi-2013a}. Finally, two coherent
states of finite amplitude are never truly
orthogonal since
$\braket{\alpha|\beta}=\exp[-(|\alpha|^2+|\beta|^2-\alpha^*\beta-\alpha\beta^*)/2]$.

Coherent states have interesting relaxation and coherence properties~\cite{Haroche-2006-book}. Being
eigenstates of the photon annihilation operator, the loss of a photon
at a rate $\kappa$ leaves a coherent state in a coherent state with damped amplitude,
i.e. $\ket{\alpha}\rightarrow\ket{\alpha e^{-\frac{\kappa}{2}t}}$. In
particular this implies that the mean photon number $\bar
n(t)=|\alpha(t)|^2$ of a coherent state decays at the single photon
loss rate $\kappa$ independently of its amplitude. In this sense a coherent state is more robust against
photon loss than a Fock state $\ket{N}$, the occupation of which
decays at a rate $N\kappa$~\cite{Scully-1969a,WangH-2008a}. However, a superposition of coherent states such as the
even parity cat state
$\mathcal{N}(\ket{\alpha}+\ket{-\alpha})$, with normalization
factor $\mathcal{N}=1/\sqrt{2+2\exp(-2|\alpha|^2)}$, decoheres at the
enhanced rate $2\bar n\kappa$ (at short times $ t\ll 1/(\bar n\kappa)$). The latter property makes it clear that quantum information
processing with cat-states will only succeed if fast and reliable
quantum error correction can be done on such qubits. Recently
encouraging results in this direction in both theory and
experiment have been obtained~\cite{Lund-2008a,Leghtas-2013b,Mirrahimi-2013a,Everitt-2013a,Sun-2013a}.

In this work, we consider specifically the following two logical qubit
encodings~\cite{Ralph-2003b}
\begin{align}\label{eq:13}
\ket{0}_L=\ket{0},\quad\ket{1}_L=\ket{\alpha}
\end{align}
and~\cite{Leghtas-2013b}
\begin{align}\label{eq:14}
\ket{\mathcal{C}_{\alpha}^+}=\mathcal{N}(\ket{\alpha}+\ket{-\alpha}),\quad \ket{\mathcal{C}_{i\alpha}^+}=\mathcal{N}(\ket{i\alpha}+\ket{-i\alpha}).
\end{align}
We shall further assume that $|\alpha|$ is sufficiently large such that
$\braket{0|\alpha}=\exp(-|\alpha|^2/2)\approx 0$ and
$\mathcal{N}\approx 1/\sqrt{2}$ (for example, for $\alpha=4$, one has
$\braket{0|\alpha}=3.4\cdot 10^{-4}$). The $\{\ket{0}_L,\ket{1}_L\}$ encoding,
which we shall call the computational encoding, turns out to be more
convenient for logical operations utilizing the dispersive interaction
between the microwaves and the transmon. The $\{\ket{\mathcal{C}_{\alpha}^+},\ket{\mathcal{C}_{i\alpha}^+}\}$
encoding, which we shall call the memory encoding, is more convenient for autonomous quantum error
correction~\cite{Leghtas-2013b}. This is due to the fact that both
$\ket{\mathcal{C}_{\alpha}^{+}}$ and $\ket{\mathcal{C}_{i\alpha}^+}$ are eigenstates with eigenvalue $+1$ of
the photon number parity operator $\bm\Pi=\exp(i\pi{\bm a}^{\dagger}{\bm a})$
and that the loss of a single photon changes the photon number parity
and can thus be detected by measuring
$\bm\Pi$~\cite{Leghtas-2013b,Sun-2013a}. In Section~\ref{sec:switch-betw-comp}, we provide a
sequence of deterministic operations to switch between these two
encodings and for the rest of this paper we focus on the
computational encoding to discuss logical operations.

\section{Synthetizing logical qubit operations\label{sec:synth-logic-qubit}}
We start by generalizing the conditional
transmon rotation (\ref{eq:3}) by allowing for simultaneous driving of multiple
photon number resolved resonances. Ideally this operation is given by
\begin{equation}\label{eq:7}
\bm X_{\theta,\phi}^{\mathcal{S}}=\sum_{n\in\mathcal{S}}\ket{n}\bra{n}\otimes
e^{i\frac{\theta}{2}\hat
  n_{\phi}\cdot\vec{\bm\sigma}}+\sum_{m\notin\mathcal{S}}\ket{m}\bra{m}\otimes\openone,
\end{equation}
where $\mathcal{S}\subset\mathbb{N}$ denotes an arbitrary set of photon
numbers. Such an operation can be approximately realized by the drive term
(\ref{eq:5}) using a pulse of duration $T_p\gg\pi/(2\mychi)$ with the
following drive frequencies, component amplitudes and phases
\begin{align}
\nu_n=\omega_{eg}-2n\mychi,\quad
\Omega_j=\delta_{jn}\frac{\theta}{T_p},\quad \delta_j=\delta_{jn}\phi,
\end{align}
where $n\in\mathcal{S}$. Imperfections will result from off-resonant
driving of undesired photon number transitions as explained in more
details in Appendix~\ref{sec:photon-number-subset}.

\begin{figure}[ht]
\begin{center}
\includegraphics[width=0.51\columnwidth]{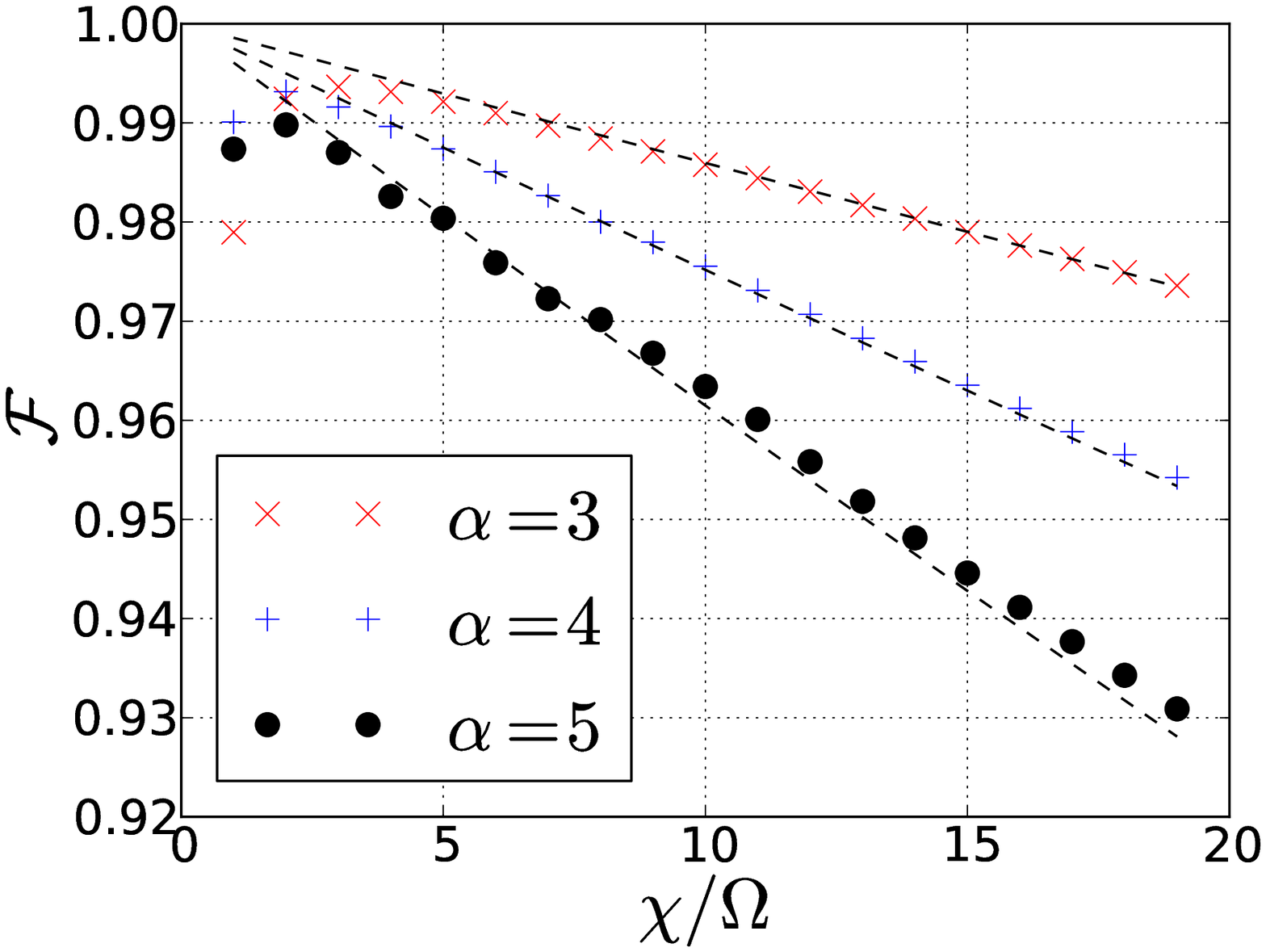}\hspace{-0.3cm}
\includegraphics[width=0.51\columnwidth]{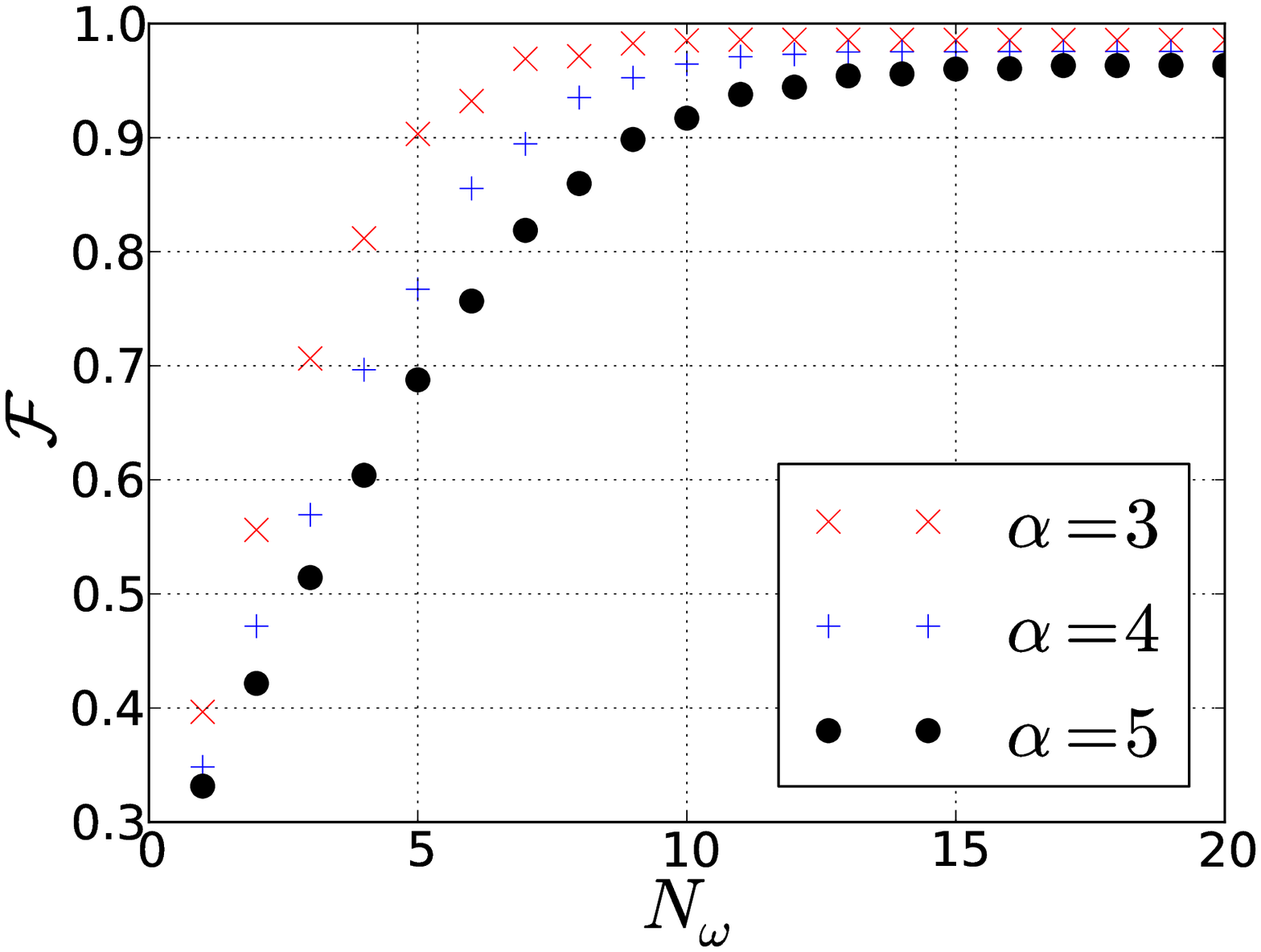}\caption{(Color
  online) Fidelity of the photon number parity conditional $\pi$-pulse from the numerical solution of Eq.~(\ref{eq:10}) with
  $\protect\mychi/(2\pi)=50\,{\rm MHz}$, $\kappa=10^{-4}\protect\mychi$. {\bf
  Left panel:} Fidelity as a function of $\protect\mychi/\Omega$. Here
$N_{\omega}=20$. Dashed
  (black) curves show $\exp[-\bar n\kappa k\pi/(2\protect\mychi)]$. {\bf Right
  panel:} Fidelity as a function of the number of drive frequency components
  $N_{\omega}$. Here $\protect\mychi/\Omega=10$. Different symbols represent
  different amplitudes $\alpha$ as given in the legend.\label{fig:2}}
\end{center}
\end{figure}

A central role in our Hadamard gate implementation is played by the
odd photon number parity conditional $\pi$-pulse given by~(\ref{eq:7}) with
$\theta=\pi$, $\phi=\pi/2$ and $\mathcal{S}=\{2n+1:\,n\in\mathbb{N}\}$. Acting on the product state
$\ket{\alpha}\otimes\ket{g}$, this gate ideally entangles the
photon number parity with the state of the transmon as follows
\begin{equation}\label{eq:8}
\bm X^{\{2n+1:\, n\in\mathbb{N}\}}_{\pi,\frac{\pi}{2}}\ket{\alpha}\otimes\ket{g}=\frac{1}{\sqrt{2}}\left(\ket{\mathcal{C}_{\alpha}^+}\ket{g}+\ket{\mathcal{C}_{\alpha}^-}\ket{e}\right),
\end{equation}
where
$\ket{\mathcal{C}_{\alpha}^-}=(\ket{\alpha}-\ket{-\alpha})/\sqrt{2}$
is the odd parity cat state. Importantly, because the occupation probability of the
Fock states in the coherent state $\ket{\alpha}$ obeys the Poisson
distribution with mean $\bar n=|\alpha|^2$ and width $\sqrt{\bar n}$,
it is sufficient to use a pulse with a finite number $N_{\omega}\gtrsim\sqrt{\bar n}$ of
frequency components $\omega_n=\omega_{eg}-2n\mychi$, distributed around $\omega_{eg}-2\bar n\mychi$. Fig.~\ref{fig:2} shows the fidelity of the simulated entangling
operation~(\ref{eq:8}) as a function of the ratio $\mychi/\Omega$
(left panel)
and as a function of $N_{\omega}$ (right panel) in the presence of
photon loss.
For square pulses, as used in the simulation, a simple pulse optimization
consists in taking $\mychi/\Omega=k\in\mathbb{N}$, since the spectral
weight of the pulse with frequency component $\omega_n$ then vanishes
identically at all other resonance frequencies $\omega_{m\not=n}$ (technically the zeros of the sinc function fall on the unwanted
photon number resonances, as explained in Appendix~\ref{sec:photon-number-subset}). As shown in the left panel of Fig.~\ref{fig:2}, the
fidelity first increases with increasing $k$, which is due to the
decreasing AC-Stark shift $\sim \Omega/k$ induced off-resonance. However, increasing $k$ also increases the
duration of the $\pi$-pulse $T_p=\pi/\Omega\sim k$ and hence increases
the effect of photon loss, which leads to the observed decrease in
fidelity. The dashed black curves on the left panel represent
$\exp[-\bar n \kappa k\pi/(2\mychi)]$ for the different coherent state
amplitudes.
On the right panel of Fig.~\ref{fig:2} we observe the expected saturation of the fidelity when the
number of frequency components becomes large compared with $\sqrt{\bar
  n}$. The parameters for the
simulation are given in the figure caption. In
Appendix~\ref{sec:photon-number-subset}, we show that similar fidelities are obtained for Gaussian
pulse envelopes. It is conceivable that more sophisticated pulse engineering could lead to
further improvement.

\begin{figure*}[ht]
\begin{center}
\includegraphics[width=2\columnwidth]{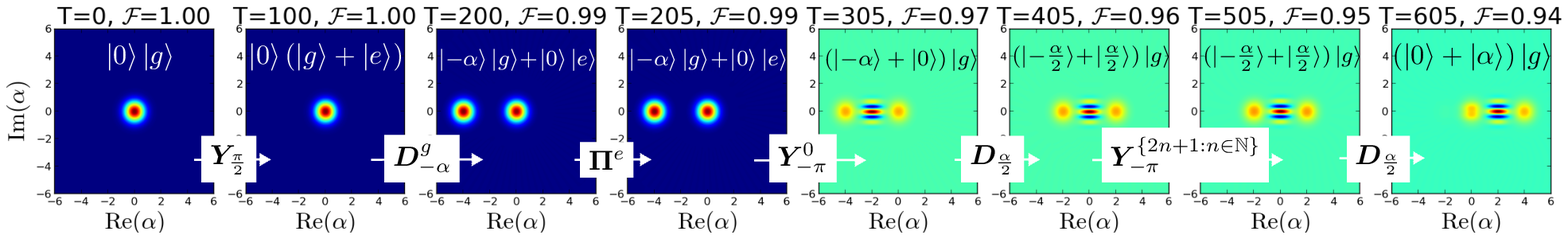}\\
\includegraphics[width=2\columnwidth]{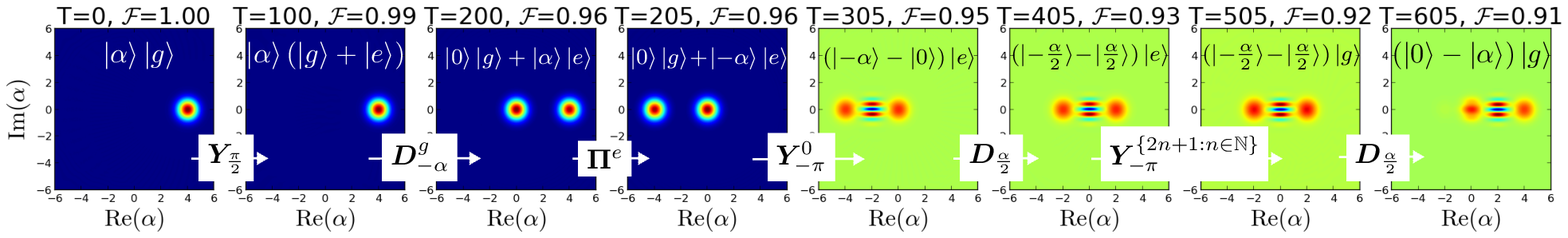}\caption{(Color
  online) Wigner
  function $W(\alpha)=(2/\pi){\rm tr}[{\bm\Pi}{\bm
  D}_{-\alpha}\rho_{\rm cav}\bm
D_{\alpha}]$ of the reduced density matrix of the cavity after each
  operation in the Hadamard gate sequence~(\ref{eq:9}). {\bf Upper panel:} for the
  initial state $\ket{0}_L$. {\bf Lower panel:} for the initial state
  $\ket{1}_L$. In this simulation we use $\alpha=4$,
  $\protect\mychi/(2\pi)=50\,{\rm MHz}$ and
  $\kappa=10^{-4}\protect\mychi$ as well as square pulses with
  duration $T_p=10\pi/\protect\mychi$ for both transmon rotations and
  field displacements. The fidelity to the target state indicated in
  each panel (omitting normalization) as well as
  the time in ${\rm ns}$ is indicated in the title of each
  sub-figure. Notice the phase difference of $\pi$ in the interference fringes
  of the two final states.\label{fig:4}}
\end{center}
\end{figure*}

\subsection{A deterministic logical Hadamard gate\label{sec:determ-logic-hadam}}
A deterministic Hadamard gate can be implemented by the following
sequence of operations (to be read from right to left)
\begin{equation}\label{eq:9}
\bm H_L=\bm D_{\frac{\alpha}{2}}\bm Y_{-\pi}^{\{2n+1:\, n\in\mathbb{N}\}}\bm
D_{\frac{\alpha}{2}}\bm Y^0_{-\pi}\bm\Pi^e\bm D^g_{-\alpha}\bm Y_{\frac{\pi}{2}},
\end{equation}
where we have used the short-hand notation
$\bm Y_{\theta}=\bm X_{\theta,\frac{\pi}{2}}$. The corresponding
circuit diagram is shown in Appendix~\ref{sec:circ-diagr-determ}.

The initial unconditional qubit
rotation $\bm Y_{\frac{\pi}{2}}$
can be realized with a short pulse with center frequency $\omega_{eg}-\bar
n\mychi$ and duration $T_p\ll 1/(2\bar n\mychi)$ such
that its frequency spectrum approximately homogeneously covers a
number of photon number resonances large compared with $\bar
n$~\cite{Leghtas-2013a}. Similarly, the unconditional cavity displacement
$\bm D_{\frac{\alpha}{2}}$ can be realized by a pulse with center frequency
$\omega_c$ and duration $T_p\ll 1/(2\mychi)$ such that the cavity resonances
corresponding to the ground and excited states of the transmon are equally
driven~\cite{Leghtas-2013a}. As discussed in Appendix~\ref{sec:beyond-two-level}, in order to
avoid spurious population of higher transmon levels, it is however
preferable to use more sophisticated pulse shaping in frequency. In
the simulations we thus use longer square pulse envelopes
($T_p=10\pi/\mychi$) with narrow frequency
bands around selected resonances for both conditional and
unconditional operations. Clearly there is a trade-off in
minimizing the adverse effects of decoherence (increasing with pulse
duration) and spurious transmon
excitations (decreasing with pulse duration) and further optimization,
beyond the scope of the present work, is desirable for an experimental
realization. The results presented below however show that even with modest
effort, rather high operation fidelities can be reached.

\begin{figure}[ht]
\begin{center}
  \includegraphics[width=0.9\columnwidth]{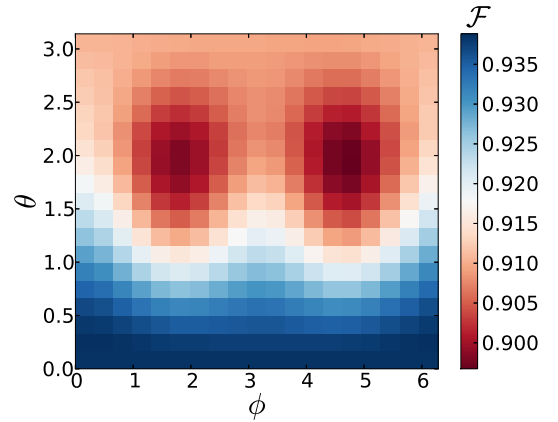}\caption{(Color
    online) Fidelity
    of the deterministic Hadamard gate as a function of the initial cavity
    state parametrized on the logical Bloch sphere. Parameters used in the simulation are
    $\protect\mychi/(2\pi)=50\,{\rm MHz}$,
    $\kappa=10^{-4}\protect\mychi$, $\alpha=4$, and square-pulse
    duration $T_p=10\pi/\protect\mychi$. The total gate duration is
    $T_{\rm tot}=6T_p+\pi/(2\protect\mychi)\approx 605\,{\rm ns}$.\label{fig:7}}
\end{center}
\end{figure}

Figure~\ref{fig:4} shows the Wigner function $W(\alpha)=(2/\pi){\rm tr}[{\bm\Pi}{\bm
  D}_{-\alpha}\rho_{\rm cav}\bm
D_{\alpha}]$ of the reduced density matrix of the cavity $\rho_{\rm
  cav}={\rm tr}_{\rm qb}[\rho]$, where the full density matrix $\rho$ is
obtained by solving the master
equation~(\ref{eq:10}) following the sequence of
operations~(\ref{eq:9}), starting with the two computational basis
states $\ket{0}_L$ (upper panel) and $\ket{1}_L$ (lower panel). The parameter values used are given in the figure
caption. The fidelity of this Hadamard gate is defined as
$\mathcal{F}=|\braket{\psi_0|\bm H\bra{g}\rho\ket{g}\bm H|\psi_0}|^2$,
where $\bm H$ is the ideal Hadamard transform. The fidelity depends on
the initial state as shown in Fig.~\ref{fig:7} where it is plotted as a function of the two angles $\theta$ and $\phi$
parametrizing the initial cavity state on the logical Bloch sphere as
$\ket{\psi_0}=\cos(\theta/2)\ket{0}_L+e^{i\phi}\sin(\theta/2)\ket{1}_L$.

\subsection{Phase gate via a conditional Berry-Phase\label{sec:phase-gate-via}}
\begin{figure}[ht]
\begin{center}
\includegraphics[width=0.8\columnwidth]{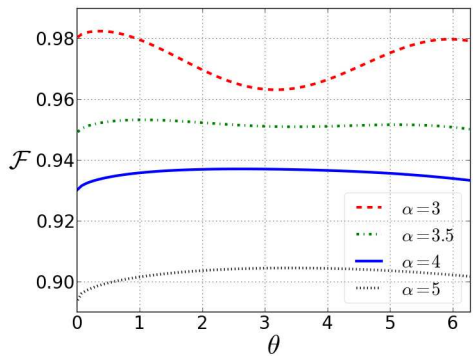}\caption{(Color
  online) Fidelity
  of the phase gate as a function of $\theta$ for
  $\protect\mychi/(2\pi)=50\,{\rm MHz}$,
  $\kappa=10^{-4}\protect\mychi$ and fixed pulse duration $T_p=10\pi/\protect\mychi$.\label{fig:12}}
\end{center}
\end{figure}
In order to achieve arbitrary single logical qubit rotations we next
explain how to implement a phase gate or arbitrary rotation around the
logical $Z$ axis. Together with the above Hadamard gate, this gives
the
ability to perform arbitrary logical $X$ rotations and therefore
arbitrary logical single qubit rotations. The phase gate is based on
the concatenation property of displacement operations:
$\bm D(\alpha)\bm D(\beta)=e^{(\alpha\beta^*-\alpha^*\beta)/2}\bm D(\alpha+\beta)$
which implies in particular the identity
\begin{equation}
\bm D_{\beta}^g\bm D_{-i\beta}^g\bm D_{-\beta}^g\bm D_{i\beta}^g=e^{2i\beta^2}\ket{g}\bra{g}+\openone\ket{e}\bra{e},
\end{equation}
where we took $\beta\in\mathbb{R}$. The phase $2\beta^2$ is in fact
nothing else but the Berry phase~\footnote{Note that for a harmonic
  oscillator the adiabaticity condition is always satisfied.} acquired by the coherent state upon
undergoing a cyclic evolution along a closed path encircling an area $\beta^2$ in
phase-space. Thus by choosing $\beta=\sqrt{\theta/2}$, the following
sequence (to be read from right to left) implements the desired phase gate
\begin{equation}\label{eq:21}
\bm Y_{\pi}^0\bm D_{\beta}^g\bm D_{-i\beta}^g\bm D_{-\beta}^g\bm
D_{i\beta}^g \bm Y_{\pi}^0=e^{i\frac{\theta}{2}}e^{i\frac{\theta}{2}\bm Z_L},
\end{equation}
where we have introduced $\bm
Z_L=\ket{1}_L\bra{1}-\ket{0}_L\bra{0}$. Figure~\ref{fig:12} shows the fidelity of
this gate as a function of the angle $\theta$ for the initial state
$\ket{\psi}_0=(\ket{0}_L+\ket{1}_L)/\sqrt{2}$ for different values of
$\alpha$. In this simulation we fix the duration of the displacement
operations. The non-monotonic dependence of the fidelity on $\theta$
is due to the interplay between the photon number dependent
decoherence rate and the $\theta$ dependent variation of the photon number during the
sequence~(\ref{eq:21}).

\subsection{Two qubit controlled phase gate\label{sec:two-qubit-controlled}}
\begin{figure}[ht]
\begin{center}
\includegraphics[width=0.9\columnwidth]{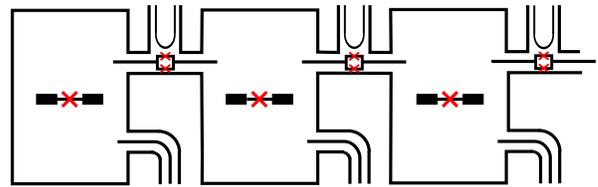}\caption{(Color
  online) Chain of
  coupled 3D-transmons. (Red) crosses represent the Josephson junctions
  at the center of each transmon. Each cavity is connected to its
  neighbors by an inline split transmon (SQUID loop). The cavities are represented by the
  rectangular boxes. The U-shaped lines between two cavities
  represents flux bias lines used to tune the
  transition frequencies of the inline-transmons. A transmission line coupling to each
  cavity further allows for coherent microwave drives.\label{fig:1}}
\end{center}
\end{figure}

Two logical qubits may be coupled by dispersively coupling the fields
of two adjacent cavities to a flux tunable split-transmon as depicted
in Fig.~\ref{fig:1}.
Assuming that the inline transmon between cavities $j$ and $j+1$ remains in its groundstate
throughout, the dispersive interaction gives rise to a photon hopping
term of the form $\xi_{jj+1}(\bm a_j^{\dagger}\bm a_{j+1}+h.c.)$. As
shown in Ref.~\onlinecite{Ralph-2003b}, such a term acting for a time $T_p$,
naturally gives rise to a controlled phase gate ${\bf
C}_{\pi}=\ket{00}\bra{00}_L+\ket{01}\bra{01}_L+\ket{10}\bra{10}_L-\ket{11}\bra{11}_L$
on the logical qubits
(assumed to have equal amplitudes $\alpha$ in the state $\ket{1}_L$)
provided that $\xi_{jj+1}T_p\bar n = \pi/2$ while $\bar
n=|\alpha|^2\gg\pi^2/4\approx 2.5$. In this limit and for the
dissipationless case, it can be shown that the worst case fidelity of this gate due to
coherent state non-orthogonality goes as
\begin{equation}\label{eq:11}
\mathcal{F}\sim\exp\left(-\frac{\pi^2}{2\bar n}\right).
\end{equation}
Since the linear decrease with $\bar n$ of the gate duration
compensates the linear increase of the effective decoherence rate ($\gamma=\bar n\kappa$), it is obviously advantageous to use a large coherent state amplitude to
increase the fidelity (\ref{eq:11}) of this gate. A suitably large
amplitude can be obtained by preceding the controlled
phase gate by the following
amplitude pumping sequence on both cavities
\begin{align}
\bm Y^0_{-\pi}\bm D_{\beta}^g\bm Y^0_{\pi}
\end{align}
with $\beta\in\mathbb{R}$ such that $\beta+\alpha>\alpha$.
After the controlled phase gate the original
amplitude can be restored by the following amplitude damping sequence on both cavities
\begin{align}
\bm Y^0_{-\pi}\bm D_{-\beta}^g\bm Y^0_{\pi}.
\end{align}
\begin{figure}[ht]
\begin{center}
\includegraphics[width=0.8\columnwidth]{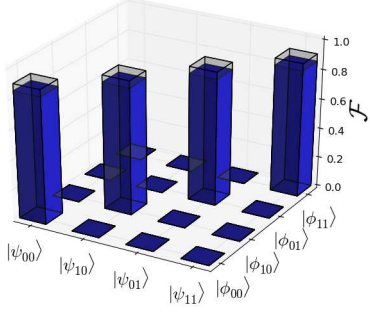}\caption{(Color
  online) Controlled
phase gate. The photon hoping strength
is $\xi/(2\pi)=25\,{\rm MHz}$ and the single photon loss rate is $\kappa=2\cdot 10^{-4}\xi$. The full bars show the results for $\alpha=4$ and the
semi-transparent bars for $\alpha=5$.\label{fig:11}}
\end{center}
\end{figure}

Figure~\ref{fig:11} shows the matrix representation of the fidelity of the simulated controlled
phase gate $\bm{C}_{\pi}$, acting on the $\bm X_L\otimes \bm X_L$ basis states
$\ket{\psi_{ij}}=(\ket{0}_L+(-1)^i\ket{1}_L)\otimes(\ket{0}_L+(-1)^j\ket{1}_L)/2$,
which are ideally mapped according to
$\ket{\psi_{ij}}\rightarrow\ket{\phi_{ij}}$ onto the entangled states
$\ket{\phi_{ij}}=(\ket{00}+(-1)^i\ket{10}+(-1)^j\ket{01}-(-1)^{i+j}\ket{11})/2$. The
fidelity is shown for two different values of $\alpha$, demonstrating the
increase in fidelity with increasing $\alpha$. Parameter
values used in the simulation
are given in the figure caption.

Together with the single qubit operations presented above, this
completes the set of universal logic gates. A similar setup, albeit
for a different two-qubit gate has been
proposed in Ref.~\onlinecite{Mirrahimi-2013a}.
\section{Switching between computational and memory encodings\label{sec:switch-betw-comp}}
\begin{figure}[ht]
\includegraphics[width=\columnwidth]{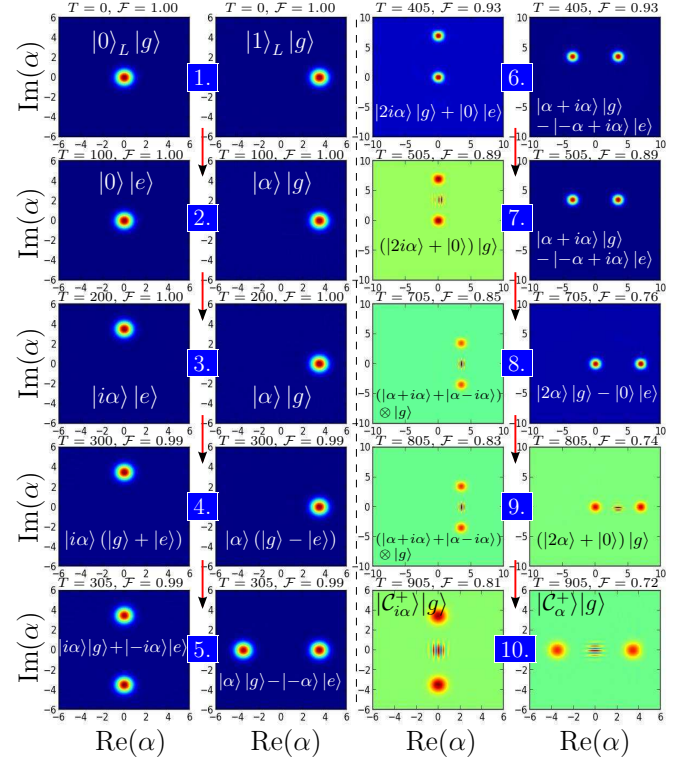}\caption{(Color
  online) Evolution of
  the Wigner function of the two logical basis states $\ket{0}_L$
  (first and third columns) and
  $\ket{1}_L$ (second and fourth columns) during the encoding-switching
  sequence~(\ref{eq:22}). Here we take $\alpha=3.5$,
  $\protect\mychi/(2\pi)=50\,{\rm MHz}$,
  $\kappa=10^{-4}\protect\mychi$ and varying pulse durations
  $T_p\in\{10\pi/\protect\mychi,40\pi/\protect\mychi\}$. In each panel
  we indicate the ideal target state. The fidelity of each step as
  well as the time in $\rm ns$ is indicated in the titles.\label{fig:13}}
\end{figure}

In~\citet{Mirrahimi-2013a} a set of universal gates for logical qubits encoded in the
memory basis (\ref{eq:14}) was given. An advantage of this proposal is that logical
operations and quantum error correction take place within the same
encoding, which allows for the logical operations to be made
fault-tolerant to a large extent~\cite{Mirrahimi-2013a}. One drawback of that
specific proposal though, is that in order to implement an arbitrary
single qubit rotation in the memory encoding it is necessary to make use of the self-Kerr
interaction among the photons $-K{\bm a}^{\dagger}{\bm
  a}^{\dagger}{\bm a}{\bm a}$ (see
Appendix~\ref{sec:infl-induc-phot}). The self-Kerr interaction
strength $K$ is
however typically orders of magnitude smaller than the dispersive
interaction~\cite{Kirchmair-2013a,Vlastakis-2013a}, leading to inappropriately long gate
times. Increasing the strength of the self-Kerr interaction is not
desirable, since it adversely affects the logical
state preparation, by squeezing the coherent states. As shown above, in the computational
encoding (\ref{eq:13}), it is possible to implement deterministically an
arbitrary single qubit rotation using the cross-Kerr term between the
photons and the transmon alone. Here
we show that the cross-Kerr term also allows to switch betweeen the two
encodings. Thus a potentially faster mode of operation would consist in keeping the logical qubits in
the protected memory encoding when idle, continuously performing error
correction as described in~\cite{Leghtas-2013b,Mirrahimi-2013a} and map them into the
computational encoding only to perform logic operations as described
in the present work. The sequence of operations to switch from the
computational to the memory encoding is given by
\begin{align}\label{eq:22}
\bm D_{-\alpha}\bm Y_{\pi}^0\bm D_{\alpha}\bm D_{-i\alpha}\bm
Y_{-\pi}^0\bm D_{i\alpha}\bm \Pi^e\bm Y_{-\frac{\pi}{2}}\bm D_{i\alpha}^e\bm Y_{\pi}^0,
\end{align}
The inverse transformation is obtained by inverting the
sequence and replacing $\alpha\rightarrow-\alpha$ as well as
$\pi\rightarrow-\pi$. Figure~\ref{fig:13} shows the evolution of the Wigner function of the logical qubit basis states after each
operation in the sequence~(\ref{eq:22}). In order to shorten the
duration of the sequence we have combined the two successive
displacements using $\bm D_{\alpha}\bm D_{-i\alpha}=e^{i\alpha^2}\bm
D_{\alpha-i\alpha}$. We find fidelities of $81\%$ and $72\%$ for the
transformations $\ket{0}_L\rightarrow\ket{\mathcal{C}_{i\alpha}^+}$
and $\ket{1}_L\rightarrow\ket{\mathcal{C}_{\alpha}^+}$ respectively.
\section{Summary \& Conclusion\label{sec:summary--conclusion}}
In summary we have proposed a deterministic Hadamard gate for logical
coherent state qubits encoded in the two-dimensional space spanned by
the vacuum and a coherent state of finite amplitude. Our scheme utilizes the
experimentally demonstrated strong dispersive
interaction of the electromagnetic field of a superconducting cavity
with an off-resonant effective two-level system (transmon qubit) and
generalizes a recently introduced set of conditional
operations~\cite{Leghtas-2013a}. We further describe a set of universal
deterministic quantum gates and provide numerical estimates for
fidelities that should be achievable in state of the art circuit QED systems.

This work was supported financially by the Swiss National Science
Foundation.

\appendix
\section{Photon-number subset conditional transmon rotation\label{sec:photon-number-subset}}
In this appendix we describe in more details how to implement
operation~(\ref{eq:7}). In a frame rotating with the bare transmon and
cavity frequencies, a photon number subset $S$ dependent transmon pulse is described
by the Hamiltonian
\begin{equation}
\bm H =-\mychi{\bm a}^{\dagger}{\bm a}{\bm\sigma}^z+\sum_{m\in
S}\frac{\Omega_m(t)}{2}\left(e^{-2im\mychi t}{\bm\sigma}^-+h.c.\right)
\end{equation} where $\Omega_m(t)$ is the envelope function of the
pulse component with center frequency $\omega_m=\omega_{eg}-2m\mychi$. Moving to the
interaction picture with the unitary transformation
\begin{equation} \bm U(t)=\exp\left[-it\mychi{\bm\sigma}^z\sum_{n}n
\ket{n}\bra{n}\right]
\end{equation} we obtain
\begin{align} \tilde {\bm H} &= \bm U\bm H\bm U^{\dagger}-i\bm U\dot
  {\bm U}^{\dagger}\\
&=\sum_{m\in
S}\frac{\Omega_m}{2}\sum_{n}\left(e^{-2i\mychi t
(m-n)}{\bm\sigma^-}+h.c.\right)\ket{n}\bra{n}\nonumber\\ &=\sum_{m\in
S}\frac{\Omega_m}{2}{\bm\sigma}^x\ket{m}\bra{m}\nonumber\\
&\quad+\sum_{m\in
S}\frac{\Omega_m}{2}\sum_{\underline n\not=m}\left(e^{-2i\mychi
t(m-n)}{\bm\sigma}^-+h.c.\right)\ket{n}\bra{n}.\nonumber
\end{align}
Consider now the perturbative expansion for $\Omega_m/\mychi\ll 1$ of the evolution
operator in this frame
\begin{align}\label{eq:15}
&\bm U_{\rm evol}(T)=\mathcal{T}\exp\left[-i\int_0^T\tilde
{\bm H}(\tau)d\tau\right]\approx 1-i\int_0^T\tilde {\bm H}(\tau)d\tau\nonumber\\
&=1-\frac{i}{2}\sum_{m\in
S}\Big\{\int_0^Td\tau\Omega_m(\tau){\bm\sigma}^x\ket{m}\bra{m}\\
&\quad+\sum_{\underline
n\not=m}\int_0^T\!\!\!\!d\tau\frac{\Omega_m(\tau)}{2}\left(e^{-2i\mychi(m-n)\tau}{\bm\sigma}^-+h.c.\right)\ket{n}\bra{n}\Big\}.\nonumber
\end{align} For $\Omega_m(\tau)=\theta(T-\tau)\Omega_0$ this becomes
\begin{align} &{\bm U}_{\rm evol}(T)\approx 1-i\frac{\Omega_0
T}{2}\sum_{m\in
S}\Big[{\bm\sigma}^x\ket{m}\bra{m}\\
&\sum_{\underline
n\not=m}e^{-i\mychi(m-n)T}\left(\frac{\sin(\mychi(m-n)T)}{\mychi(m-n)T}{\bm\sigma}^-+h.c.\right)\ket{n}\bra{n}\Big].\nonumber
\end{align} Now let us choose $T$ such that $\mychi T=\pi$. Then the
second term above vanishes identically, as the zeros of the sinc function line up with the photon resonances of the
transmon that we do not wish to drive and we have
\begin{align}
{\bm U}_{\rm evol}(\pi /\mychi)&\approx 1-i\pi\left(\frac{\Omega_0 }{2\mychi}\right)\sum_{m\in
S}{\bm\sigma}^x\ket{m}\bra{m}.
\end{align} This generates a small $x$-rotation of the transmon
components associated with the Fock states $\ket{m}$ with $m\in S$ by an angle $\pi
\Omega_0/\mychi\ll\pi$. By concatenating such
rotations $k$ times, effectively increasing the total pulse time to
$T=k\pi/\mychi$, we can then generate a rotation of the transmon by an
angle $k\pi\Omega_0/\mychi$ as
\begin{align} &{\bm U}_{\rm evol}(k\pi/\mychi)\approx {\bm U}_{\rm
evol}(\pi/\mychi)^k\\
&=\exp\left[-i\frac{k\pi\Omega_0}{2\mychi}{\bm\sigma}^x\right]\sum_{m\in
S}\ket{m}\bra{m}+\sum_{n\notin S}\ket{n}\bra{n}.\nonumber
\end{align}
By changing the phase of the drive signal we may similarly
implement a photon parity conditional rotation around arbitrary axes
in the $x-y$ plane. For example we may perform a $\pi$ rotation around
the $y$-axis conditioned on the parity of the photon number being odd
as in the main text by selecting a drive phase of $\pi$, $S=\{2n+1\}_{n\in\mathbb{N}}$ and
$k\Omega_0=\mychi$, which implies $k\gg 1$ given that we assumed
$\Omega_0\ll\mychi$. Higher order terms in the perturbative
expansion~(\ref{eq:15}), lead to correction terms of order
$\Omega_0^2/\mychi$, which correspond to the
AC-stark shift~\cite{Vlastakis-2013a}.
\begin{figure}[ht]
\begin{center}
\includegraphics[width=0.8\columnwidth]{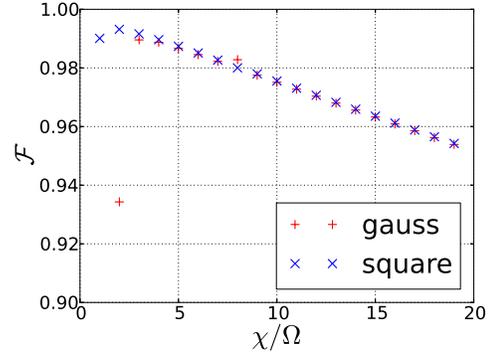}\caption{(Color
  online) Comparison
  of the fidelities of the photon number parity conditional
  $\pi$-pulse between gaussian and square pulses. For small
  $\protect\mychi/\Omega$, the fidelity is lower for the gaussian
  pulses which is due to the larger on-resonance driving of unwanted
  transistions in this case. For larger $\protect\mychi/\Omega$, the obtained
  fidelities are similar. We take $\alpha=4$, $N_{\omega}=20$,
  $\protect\mychi/(2\pi)=50\,{\rm MHz}$ and $\kappa=10^{-4}\protect\mychi$.\label{fig:5}}
\end{center}
\end{figure}

In reality
one cannot generate perfect square pulses and this will lead to
corrections. It is easily seen that for gaussian pulse envelopes, the
strength of the
residual on-resonance drives of other resonances is suppressed at least by the
super-exponential factor
$\sum_{k=1}^{\infty}e^{-(2\mychi Tk)^2}$ and thus becomes negligible
for a pulse duration $T\gg 1/(2\mychi)$. Note also that typically for
a conditional qubit rotation by an angle $\theta\in[0,2\pi]$, $T\sim \theta/\Omega_0$ which means the
correction terms will be small as long as $\Omega_0\ll\mychi$
consistent with our assumption. Figure~\ref{fig:5} shows the fidelity of the odd
photon number parity conditional $\pi$-pulse implemented with gaussian
pulses $A_{\pi}\exp[-(\sigma (t-T/2))^2/2]$ with frequency width $\sigma=6/T$,
total pulse duration $T_p=10\pi/\mychi$ and
amplitude $A_{\pi}=\pi\sigma/\sqrt{2\pi}$. The comparison with the
square pulse case, shows that similar fidelities can be reached in
both cases.

\section{NOT gate\label{sec:not-gate}}
Although one may in principle implement any single qubit gate with the
operations described in the main text, the dispersive coupling may
allow for more efficient direct implementations of certain gates. For
example, a logical quantum NOT gate
$\bm X_L$ can be implemented solely with unconditional operations by the sequence of
operations
\begin{equation}
\bm X_L=\bm D_{\frac{\alpha}{2}}\bm Y_{\pi}\bm \Pi^e\bm Y_{\pi}\bm D_{-\frac{\alpha}{2}}.
\end{equation}

\section{Circuit diagram of the deterministic Hadamard gate\label{sec:circ-diagr-determ}}
\begin{figure}[ht]
\begin{center}
\includegraphics[width=0.8\columnwidth]{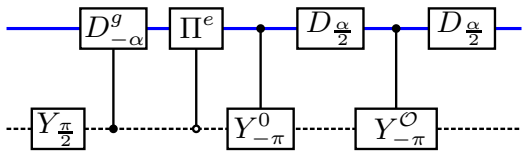}\caption{(Color
  online) Circuit
diagram for the deterministic Hadamard gate. The horizontal
full (blue) line represents the cavity state (logical qubit), while the
dashed (black) line represents the transmon state. Here $\mathcal{O}=\{2n+1:\,n\in\mathbb{N}\}$.\label{fig:3}}
\end{center}
\end{figure}

\section{Influence of the induced photon-photon interaction\label{sec:infl-induc-phot}}
In the above discussion we have neglected the self-Kerr interaction
between the photons induced by the coupling to the nonlinear transmon
oscillator~\cite{Bourassa-2012b,Nigg-2012a,Kirchmair-2013a}. In this appendix, we show that the leading order
effect of the self-Kerr can be accounted and corrected for to a large
extent. The self-Kerr effect is described by the term
\begin{equation}\label{eq:16}
\bm H_K=-K{\bm a}^{\dagger}{\bm a}^{\dagger}{\bm a}{\bm a}.
\end{equation}
Note that this term commutes with the photon number operator and is independent
of the state of the transmon and thus commutes with the conditional
transmon operations~(\ref{eq:3}) and (\ref{eq:7}). When acting on a coherent state
$\ket{\alpha}$ the leading order effect is a phase rotation~\cite{Walls_Milburn-2008}
\begin{equation}
e^{-i\bm H_K t}\ket{\alpha}\approx\ket{\alpha e^{i\phi_Kt}}\,\quad
Kt\ll 1
\end{equation}
with $\phi_K=2\bar n K$. Sub-leading order terms describe
squeezing~\cite{Walls_Milburn-2008}. To leading order, the effect of
the self-Kerr on the photon number parity operation~(\ref{eq:4}) can
thus be corrected for simply by shortening the waiting time according to
\begin{align}\label{eq:17}
T_{\pi,\bar n}=\frac{\pi}{2(\mychi+K\bar n)},
\end{align}
for a coherent state with average photon number $\bar n$. The
self-Kerr term~(\ref{eq:16}) does not commute with the displacement
operation~(\ref{eq:2}). Let us consider the case where the transmon is in
the ground state. In the presence of an on-resonance drive with constant strength
$\varepsilon\in\mathbb{R}$ and of duration $T$ such that $KT\ll 1$, the Heisenberg
equation of motion for the annihilation operator, neglecting squeezing
terms reads
\begin{align}\label{eq:20}
\dot {\bm a}(t)=2iK(\varepsilon t)^2{\bm a}+\varepsilon.
\end{align}
The solution of Eq.~(\ref{eq:20}) for the coherent state amplitude is
\begin{align}
a(t) = \left(\varepsilon\int_0^td\tau e^{-i\frac{2}{3}K\varepsilon^2\tau^3}+a(0)\right)e^{i\frac{2}{3}K\varepsilon^2t^3}.
\end{align}
Specifically for the case where the field is initially the vacuum
($a(0)=0$) and for $(\varepsilon t)^2Kt\ll 1$ we find
\begin{align}\label{eq:19}
a(t)\approx \varepsilon t+i\frac{K\varepsilon^3t^4}{2}.
\end{align}
The self-Kerr induced phase rotation during the displacement is thus
\begin{equation}\label{eq:18}
\phi_K(t)\approx\tan\phi_K=\frac{{\rm Im}[a(t)]}{{\rm
    Re}[a(t)]}=\frac{K\bar n(t) t}{2},
\end{equation}
where we have defined $\bar n(t)=(\varepsilon t)^2$. \begin{figure}[ht]
\begin{center}
\includegraphics[width=\columnwidth]{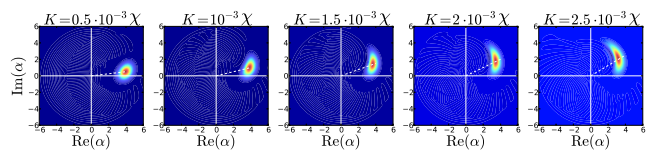}\caption{(Color
  online) Effect of
  the self-Kerr interaction on the displacement operation. The displacement amplitude
  is $\alpha=4$ and the square pulse duration is
  $T_p=10\pi/\protect\mychi$. We further take
  $\protect\mychi/(2\pi)=50\,{\rm MHz}$ and $\kappa=0$. The initial state is
  $\ket{0}_L\ket{g}$ and the Wigner function is calculated by solving numerically
  the master equation~(\ref{eq:10}) with Hamiltonian ${\bm
    H}=i\varepsilon({\bm a}-{\bm a}^{\dagger})-K{\bm a}^{\dagger}{\bm a}^{\dagger}{\bm a}{\bm a}$. The
  dashed (white) lines show $a(T_p)$ after Eq.~(\ref{eq:19}).\label{fig:8}}
\end{center}
\end{figure}
Figure~\ref{fig:8} compares Eq.~(\ref{eq:19}) for different values of
$K$ with the result of a Wigner function computation including
higher order terms. As we can see, the direction of the center of the
distribution is well captured by~(\ref{eq:19}) (dashed white lines) even in the presence of
sizable squeezing. To emphasize the effect of the self-Kerr we set
$\kappa=0$ in this simulation.

\begin{figure*}[ht]
\begin{center}
\includegraphics[width=2\columnwidth]{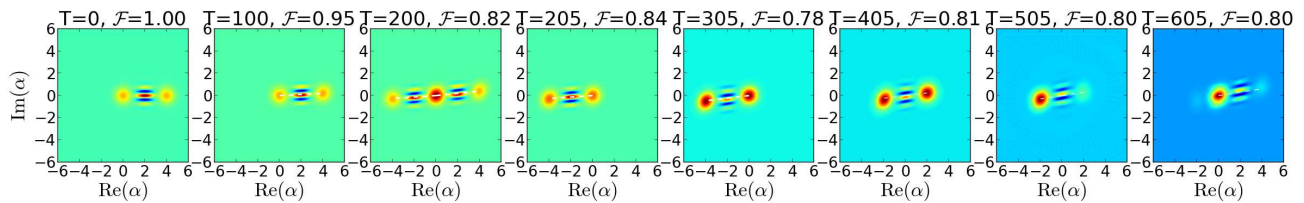}\caption{(Color
  online) Effect of
  the self-Kerr interaction on the Hadamard gate. We use a
  displacement of amplitude
  $\alpha=4$ and pulse duration $T_p=10\pi/\protect\mychi$ and take
  $\protect\mychi/(2\pi)=50\,{\rm MHz}$ and $\kappa=10^{-4}\protect\mychi$. The initial state is
  $(\ket{0}_L+\ket{1}_L)/\sqrt{2}$ and the Wigner function is calculated by solving numerically
  the master equation~(\ref{eq:10}) including the self-Kerr
  term~(\ref{eq:16}) with $K=0.5\cdot 10^{-4}\protect\mychi$.\label{fig:9}}
\end{center}
\end{figure*}

Figure~\ref{fig:9} shows the cavity Wigner function during the Hadamard
gate sequence acting on the initial state $(\ket{0}_L+\ket{1}_L)/\sqrt{2}$, including a self-Kerr term of strength $K=0.5\cdot 10^{-4}\mychi$. Here we include photon loss as
well. In the simulation we have adjusted the
displacement phases and amplitudes according to Eq.~(\ref{eq:19}) and
corrected the wait time according to Eq.~(\ref{eq:17}). The results
demonstrate that high fidelities of about $80\%$ can be reached
even in the presence of a weak self-Kerr term.

\section{Beyond the two level approximation -- conditional vs
  unconditional operations\label{sec:beyond-two-level}}
Above, we have neglected the higher excited states of
the transmon. However, a transmon is best viewed as a weakly
anharmonic oscillator with negative anharmonicity~\cite{Koch-2007a,Nigg-2012a}. Denoting the transmon
eigenstates with $\ket{j}$, the dispersive Hamiltonian in the
multi-level case reads
\begin{equation}
\sum_{j=0}\mychi_{j,j+1}\left(\ket{j+1}\bra{j+1}-\ket{j}\bra{j}\right){\bm
  a}^{\dagger}{\bm a}.
\end{equation}
Thus the $e\leftrightarrow g$ and $f\leftrightarrow e$ transitions split up into photon-number
resolved ladders of resonances according to
\begin{equation}
\omega_{n}^{(eg)}=\omega_{eg}-2n\mychi
\end{equation}
and
\begin{equation}
\omega_{n}^{(fe)}=\omega_{fe}-2n\mychi^{\prime}
\end{equation}
with $\mychi=\mychi_{01}-\mychi_{12}/2$ and $\mychi^{\prime}=\mychi_{12}-(\mychi_{23}+\mychi_{01})/2$.

\begin{figure}[ht]
\begin{center}
\includegraphics[width=0.8\columnwidth]{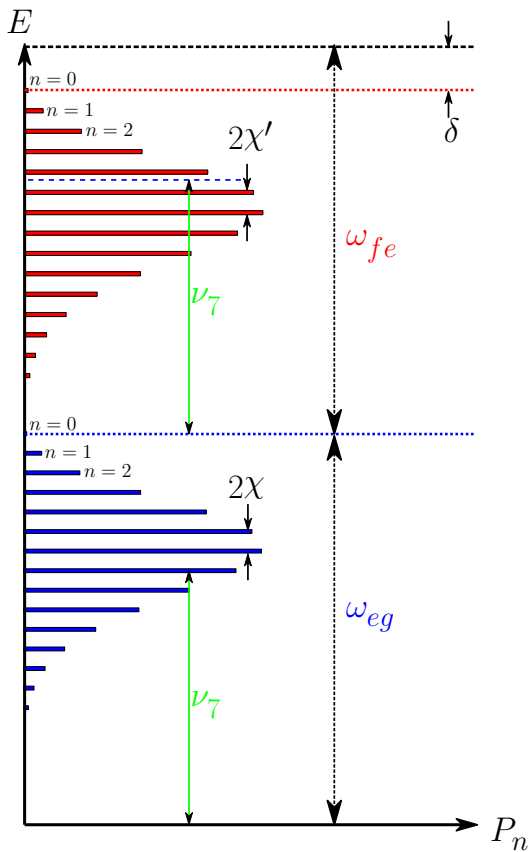}\caption{(Color
  online) Illustration
  of the 3-level transmon spectrum in the dispersive coupling regime
  with a coherent state in the cavity. Note that
  $\protect\mychi\not=\protect\mychi'$. Thus, by using
  sufficiently narrow pulse envelopes in frequency, one can avoid
  population of the excited state $f$. This is illustrated on the
  figure for the $n=7$ transition.\label{fig:10}}
\end{center}
\end{figure}

If the magnitude $\delta$ of the anharmonicity were much
larger than $2\bar n\mychi$ where $\bar n$ is the average photon
number in the logical qubit state $\ket{1}_L$, then one could
implement unconditional operations on the transmon simply by using
pulses of short duration $T$ such that $\delta\gg 1/T\gg 2\bar
n\mychi$. If such a wide separation of frequency scales is not
available, as is the case in current experimental
realizations, then it may be preferable to implement the
unconditional operations in a similar way to the conditional
operations, by superposing narrow frequency band pulses to avoid spurious population of higher
transmon levels (see Fig.~\ref{fig:10}). The penalty for this is an
increase in gate duration.

\bibliography{/home/sen/phys/library/bibtex/mybib}

\end{document}